%% file: ms.tex
\documentclass[a4paper,11pt]{article}
\pdfoutput=1 

\usepackage{jcappub} 

\usepackage[T1]{fontenc} 
\usepackage{subcaption}
\usepackage{physics}

\title{\boldmath Effects of Neutrino Masses and Asymmetries on Dark Matter Halo Assembly}

\author[a]{Hiu Wing Wong}
\author[a]{and Ming-chung Chu}

\affiliation[a]{Department of Physics, The Chinese University of Hong Kong, Shatin NT, Hong Kong}

\emailAdd{hwwong@phy.cuhk.edu.hk}
\emailAdd{mcchu@phy.cuhk.edu.hk}

\abstract{Massive cosmological neutrinos suppress the Large-Scale Structure (LSS) in the Universe by smoothing the cosmic over-densities, and hence structure formation is delayed relative to that in the standard Lambda-Cold Dark Matter ($\Lambda$CDM) model.
We characterize the merger and mass accretion history of dark matter halos with the halo formation time $a_{1/2}$, tree entropy $s$ and halo leaf function $\ell(X)$ and measure them using neutrino-involved N-body simulations. 
We show that a non-zero sum of neutrino masses $M_\nu$ delays the $a_{1/2}$ for halos with virial mass between $10^{13} M_\odot$ and $3\times 10^{13} M_\odot$, whereas a non-zero neutrino asymmetry parameter $\eta^2$ has the opposite effect.
While the mean tree entropy $\bar s$ does not depend significantly on either $M_\nu$ or $\eta^2$, the halo leaf function does.  Furthermore, the dependencies of $\ell$ on $M_\nu$ and $\eta^2$ have significant evolution in redshift $z$, with the relative contributions of $M_\nu$ and $\eta^2$ showing a sigmoid-like transition as a function of $z$ around $z \approx 0.6$.
Together with the matter power spectrum, these halo parameters allow us to break the parameter degeneracy between $M_\nu$ and $\eta^2$ so that they can both be constrained in principle.
}

\begin{document}
\maketitle
\flushbottom

\input{Maintext/Introduction}

\input{Maintext/grid-based}
\input{Maintext/halo-stat}
\input{Maintext/result}

\input{Maintext/conclusion}

\acknowledgments
We acknowledge Shek Yeung for modifying \texttt{CosmoMC} to fit the Planck 2018 data with different neutrino cosmologies and performing all the MCMC refittings. 
HWW thanks Jianxiong Chen for his mentorship on N-body simulations and Zhichao Zeng for discussion on neutrino-involved simulations.
All simulations were performed using the Central Research Computing Cluster at CUHK.
This work is supported partially by grants from the Research Grants Council of the Hong Kong Special Administrative Region, China (Project Nos. AoE/P-404/18 and C7015-19G).

\appendix
\input{Maintext/linear-evo}
\input{Maintext/posterior}


\end{document}

%% file: Maintext/Introduction.tex
\section{Introduction}
\label{sec:intro}
There are still many open questions about neutrinos, such as what their masses are, and whether they are Dirac or Majorana particles. 
Neutrino flavor oscillation experiments show that at least two types of neutrinos must be massive. Assuming normal hierarchy and the smallest neutrino mass eigenvalue to be zero, (i.e. $m_{\nu_3} >> m_{\nu_2} > m_{\nu_1}=0$, $m_{\nu_i}$, $i=1,2,3$ denoting mass eigenstates), we can put a lower bound on $M_\nu \equiv \sum_i m_{\nu_i} > 0.06$ eV \cite{PP}.

Particle physics experiments such as neutrinoless double beta decay measurements and neutrino mass spectrum experiments also try to answer these open questions. If neutrinos are Majorana particles, neutrinoless double beta decays are possible.
However, there is no solid evidence for such decay events yet. Since the mass of neutrinos is related to the decay rate, the non-observation of neutrinoless double beta decay suggests an upper bound on the effective Majorana mass of electron-type neutrinos $m_{ee}=\abs{\sum_i U_{ei}^2 m_{\nu_i}}<0.34$ eV (90\% C.L.) \cite{FARZAN200159}.

Neutrino mass spectrum experiments, on the other hand, measure the endpoint of the electron energy spectrum in beta decays. The upper bound on the mass of anti-electron neutrinos $m_{\Bar{\nu}_e} < 2.5$ eV (95\% C.L.) was obtained from Troitzk's results \cite{PP}. 
This bound from direct detection is quite conservative, and it is only valid for electron-type neutrinos. 
We know nothing about muon and tau neutrinos from these experiments.

The Cosmic Neutrino Background (C$\nu$B) has long been studied, but direct detection of the C$\nu$B is very difficult since neutrinos only participate in weak and gravitational interactions. Currently, the tightest upper bound on $M_\nu$ is provided by Planck from Cosmic Microwave Background (CMB) Anisotropies data. 
A non-zero $M_\nu$ will delay the radiation-matter equality epoch and modify the Hubble expansion rate which in turn impacts the CMB power spectrum. Planck 2018 gives a constraint on $M_\nu<0.12$ eV (95\% C.L.) \cite{Planck2018,M nu}.

Besides CMB data, the Large-Scale Structure (LSS) in the Universe is another powerful tool to study neutrino cosmology, as the LSS is more sensitive to the value of $M_\nu$ than CMB.
The structure growth is governed by three competing factors: the expansion of the universe, the kinetic energy and self-gravitation of matter, and massive neutrinos play a role in all of them. 
In the standard $\Lambda$CDM model, neutrinos are treated as radiation, whereas massive neutrinos transform from being ultra-relativistic (radiation-like) to non-relativistic (matter-like) as the universe expands and cools, resulting in a small but non-negligible change in the expansion history compared to that of $\Lambda$CDM. 
Cosmological neutrinos are considered as Hot Dark Matter (HDM) with large thermal velocities. They do not cluster significantly on small scale and tend to erase structures below the free-stream scale $k_{fs}$. This neutrino free-streaming effect is well studied in the linear regime using Boltzmann codes such as \texttt{CAMB} \cite{camb} and \texttt{CLASS} \cite {class}. 
However, the linear method breaks down when the density contrasts exceed unity, such as in a dark matter halo. In the non-linear regime, the N-body method should be used. Different methods have been proposed to incorporate neutrino effects into N-body simulations, such as the particle-based \cite{part}, grid-based \cite{grid, grid2, Carton}, "SuperEasy" \cite{Yvvone} and fluid-based \cite{Yvvone fluid, nuconcept} methods. 
They give consistent results, showing that massive neutrinos suppress the matter power spectrum below their free-streaming scale ($k>>k_{fs}$) by up to $\mathcal{O}(5\%)$ for $M_\nu = 0.06$ eV compared to that without the free-streaming neutrinos, and such a difference would be measurable with modern observation programs.

Another property of neutrinos, which governs the asymmetries of neutrinos and anti-neutrinos, is the neutrino chemical potentials $\{\mu_i\}$. The chemical potentials of anti-neutrinos would be $\{-\mu_i\}$. If neutrinos are Majorana particles, $\mu_i=0$.
As the neutrino distribution functions are frozen after decoupling, the dimensionless quantities $\{\xi_i=\mu_i/k_BT\}$ are constant throughout the expansion of the universe, where $k_B$ and $T$ are the Boltzmann constant and neutrino temperature, respectively.
We also know from Big Bang Nucleosynthesis (BBN) that the chemical potential for electron-type neutrinos is very small.
We follow \cite{Carton} and \cite{mu} to set $\xi_e=0$ and $\xi_\mu=\xi_\tau$, so that we have only one independent parameter, $\eta^2 = \sum_i \xi_i^2$, denoted the neutrino asymmetry parameter. 
The fact that the muon and tau neutrinos have strong mixing, as shown in neutrino oscillation experiments, makes $\xi_\mu=\xi_\tau$ a good approximation \cite{bbn1, bbn2}. Currently, the total neutrino asymmetry $\sum_i \xi_i$ is tightly constrained by the BBN Helium-4 mass fraction $Y_p$. However, $Y_p$ itself cannot constrain $\eta^2$, particularly since $\xi_1$ and $\xi_2$ tend to have opposite signs and cancel each other quite well \cite{eta_effect}. $\eta^2$ is only mildly constrained by other observational data such as the CMB power spectrum as we will discuss below.

When neutrino masses and asymmetries are considered, the cosmological parameters obtained from fitting of CMB data would be different from those of $\Lambda$CDM \cite{nu CMB}, and any change in the cosmology will then affect the LSS formation. To ensure self-consistency, refitting of cosmological parameters is needed for each $M_\nu$ and $\eta^2$, using a Markov-Chain Monte-Carlo (MCMC) code such as \texttt{CosmoMC} \cite{cosmomc}. In Planck 2018, the standard cosmological parameters are obtained by assuming three neutrino species, two massless states $(m_{\nu_1} = m_{\nu_2} = 0)$ plus a single massive neutrino of mass $m_{\nu_3} = 0.06$ eV, without any neutrino asymmetries \cite{Planck2018}. Therefore, we choose a baseline of $M_\nu = 0.06$ eV and $\eta^2 = 0$ to compare against when analyzing the simulation results.

The authors in \cite{Carton} explicitly tested the effect of free-streaming neutrinos with chemical potential on the matter power spectrum, and the results in Figure 3a in \cite{Carton} show that the free-streaming effect of neutrinos is not affected by $\eta^2$.
Rather, the effect of $\eta^2$ on the structure growth is mainly due to the changes in the refitted cosmological parameters and therefore the expansion history of the universe.
In \cite{eta_effect}, $M_\nu$ and $\eta^2$ are treated as free parameters and varied together with other $\Lambda$CDM parameters to fit the Planck CMB data. The finding is that $M_\nu$ and $\eta^2$ work against each other in virtually every cosmological parameter.
For instance, while $M_\nu$ has a negative correlation with both $H_0$ and $\sigma_8$, $\eta^2$ has a positive correlation with both (see Figure 10 in \cite{eta_effect}). 
We would then expect $M_\nu$ and $\eta^2$ to have the opposite effects on the structure growth via their effects on the cosmological parameters.

Indeed, a recent study showed that there is a parameter degeneracy between $M_\nu$ and $\eta^2$ in their effects on the matter power spectrum, as a non-zero $\eta^2$ would enhance the matter power spectrum and compensate the suppression from a finite $M_\nu$ \cite{Carton}. To break such a degeneracy, other cosmological observables should be considered. Since massive neutrinos suppress large-scale structures, it's natural to ask how neutrinos alter the merging and assembling of dark matter halos, as these processes are highly non-linear and very sensitive to the initial conditions of halo formation. 

There are two different ways to look at the halo assembly history: the mass accretion history and the halo merger history. The latter, characterized by the halo merger tree, keeps track of how smaller halos merge to become a bigger halo, which is different from the concept of the mass growth rate. The mass accretion history is easily quantified by the time needed for a particular halo to double its mass, which is the traditional definition of the halo formation time $a_{1/2}$ \cite{formtime}. 

Recently, the idea of the tree entropy $s$ was proposed \cite{tree entropy}, which is based on Shannon's information entropy. The tree entropy $s$ captures the complexity and geometry of a halo merger tree. The tree entropy $s$ is shown to be information-rich and especially useful for linking the galaxies to their host halos. For example, the morphology of a galaxy is closely related to the merger history of its host halo; many semi-analytical models use the merger tree of the galaxy's host halo to predict its morphology. 
A clear positive correlation between the tree entropy $s$ and the galaxy's bulge-to-total mass ratio was found in the mock galaxy catalog generated by a semi-analytical model \cite{tree entropy}. If neutrinos indeed bring a significant impact to $s$, we may be able to constrain $M_\nu$ and $\eta^2$ by measuring the morphologies of galaxies.

Finally, the halo leaf function $\ell(X)$ is defined to be the number of halos with more than $X$ leaves in their merger tree. Such a measure is conceptually similar to the halo mass function, but we bin the halos according to their merger histories instead of their masses. 

In this work, we investigate the effects of the sum of neutrino masses $M_\nu$ and neutrino asymmetry parameter $\eta^2$ on the halo assembly history, by studying the halo formation time $a_{1/2}$, tree entropy $s$ and halo leaf function $\ell(X)$. These parameters together with the matter power spectrum will allow us to break the parameter degeneracy between $M_\nu$ and $\eta^2$.

This paper is organised as follows. In Section \ref{sec:grid} we briefly introduce the grid-based neutrino method in N-body simulations along with the simulation parameters. The halo assembly statistics are elaborated in Section \ref{sec:halo}. The simulation results are examined in Section \ref{sec:result}, where we present an empirical formula for the neutrino effects on the halo assembly statistics. The discussion and conclusion are in Section \ref{sec:conclusion}.

%% file: Maintext/grid-based.tex
\section{Grid-based neutrino simulation}
\label{sec:grid}

To study the neutrino free-streaming effect on LSS, we can include neutrinos as another type of particles in N-body simulations. This particle-based method is accurate in principle but computationally expensive.
Not only will it bring extra particles and interactions, but it also requires more integration steps compared to the Cold Dark Matter (CDM)-only simulation with the same number of particles due to the larger velocity dispersion of the neutrinos. 
On the other hand, the grid-based neutrino-involved simulation includes only CDM particles in the simulation box. The neutrino information is carried by the neutrino over-density field $\delta_\nu$ contained in the Particle-Mesh (PM) grid, which is responsible for the long-range interaction in a Tree-PM code like \texttt{Gadget2} \cite{Gadget2}. 

Another advantage of the grid-based method is that we can also investigate the effects of the chemical potentials of neutrinos, which can be incorporated easily in the simulation through the Fermi-Dirac distribution of the cosmological neutrinos.

\subsection{Linear evolution for neutrino over-density}
The linear equation that governs the evolution of $\delta_\nu$ is \cite{LBE}: 
\begin{equation}
\label{eq:LBE}
    \Tilde{\delta}_\nu (\chi,\mathbf{k}) = \Phi(\mathbf{k}\chi) \Tilde{\delta}_\nu(0,\mathbf{k}) +
    4\pi G \int^\chi_0 a^4(\chi')(\chi-\chi') \Phi[\mathbf{k}(\chi-\chi')]
    [\Bar{\rho}_{cdm}(\chi') \Tilde{\delta}_{cdm} (\chi',\mathbf{k}) + \Bar{\rho}_\nu(\chi') \Tilde{\delta}_\nu (\chi',\mathbf{k})] d\chi'.
\end{equation}

Here, $\Tilde{\delta}_\nu$ ($\Tilde{\delta}_{cdm}$) and $\bar \rho_\nu$  $(\bar\rho_{cdm})$ are the over-density in Fourier space and mean density of neutrinos (CDM), respectively. $\bf k$ is the wave vector, $\chi$ is the co-moving coordinate where $d\chi=dt/a^2(t)$, and $\Phi$ is a special function that will be discussed in Appendix \ref{app:phi}.

Neutrinos cannot cluster below their free-streaming scale, which is larger than their non-linear scales; therefore, the evolution of neutrinos is well described by the linear equation. Although we use a linear equation to describe the evolution of $\delta_\nu$, the non-linear $\delta_{cdm}$ (from N-body simulation) is involved in the evolution of neutrinos. Hence, the non-linearities in structure formation are still fully preserved. 
Previous studies have also shown that both particle-based and grid-based simulations produce consistent results for the matter power spectrum \cite{Carton}.

\subsection{Total over-density}
Once we obtain the neutrino over-density $\delta_\nu$, the total over-density field $\delta_t$ is then given by:
\begin{equation}
\label{eq:avg}
    \delta_t = (1-f_\nu)\delta_{cdm} + f_\nu \delta_\nu,
\end{equation}
where $f_\nu=\Omega_\nu / \Omega_m$, the ratio of the cosmological neutrino and CDM densities. Although $f_\nu$ is small, the non-linearity in structure formation will mix up different modes, and the final density $\delta_t $ may change by a factor much greater than $(1- f_\nu)$.

\subsection{Implementation}
We implemented the grid-based neutrino method in our own modified version of \texttt{Gadget2}. The detailed procedure is discussed in \cite{Carton}. Here we briefly summarize the steps:

\begin{enumerate}
    \item The initial snapshot generated by \texttt{MUSIC} \cite{music} and initial neutrino power spectrum generated by \texttt{CAMB} are fed into \texttt{Gadget2} as the initial conditions, and thus we have both $\Tilde{\delta}_\nu(0,k)$ and $\Tilde{\delta}_{cdm}(0,k)$.
    
    \item To evolve the system, the CDM particles are drifted first. With a new CDM power spectrum after the drift $\Tilde{\delta}_{cdm}(\chi, k)$, we solve Eq.(\ref{eq:LBE}) iteratively. We use linear interpolation to approximate $\Tilde{\delta}_{cdm}(\chi', k)$ for $\chi' \in [0,\chi]$ as the time difference is usually small between two PM steps.
    
    \item The over-densities are assumed to carry the same phase:
    \begin{equation}
        \Tilde{\delta}_\nu (\chi, {\bf k}) = \frac{\Tilde{\delta}_\nu(\chi, k)}{\Tilde{\delta}_{cdm}(\chi, k)} \Tilde{\delta}_{cdm}(\chi, {\bf k}).
    \end{equation}
    The total over-density field $\Tilde{\delta}_t(\chi, {\bf k})$ is then obtained using Eq.(\ref{eq:avg}). Finally the original $\Tilde{\delta}_{cdm}(\chi, {\bf k})$ is replaced by $\Tilde{\delta}_t(\chi, {\bf k})$ to give the correct PM potential to evolve the CDM particles.
    
    \item $\Tilde{\delta}_\nu(\chi, k)$ and $\Tilde{\delta}_{cdm}(\chi, k)$ are stored as the initial conditions for the next PM calculation, and we iterate back to step 1 until the final time (usually today).
    
\end{enumerate}

\subsection{Simulation parameters}

To specify the fiducial cosmology for our N-body simulation, 6 cosmological parameters are needed. 
They are the physical CDM density $\Omega_ch^2$, the physical baryon density $\Omega_bh^2$, the observed angular size of the sound horizon at recombination $\theta$, the reionization optical depth $\tau$, 
the initial super-horizon amplitude of curvature perturbations $A_s$ at $k$ = 0.05 Mpc$^{-1}$ and the primordial spectral index $n_s$. 
All of them are consistently refitted from the Planck CMB data using the Planck 2018 \texttt{plikHM\_TTTEEE} likelihood for each set of $M_\nu$ and $\eta^2$ values. The parameters relevant to the N-body simulations are listed in Table \ref{tab:simparam}, and their posterior distributions for selected sets of $M_\nu$ and $\eta^2$ are plotted in Figure \ref{fig:post}.

\begin{table}[!hbt]
		\begin{center}
		\begin{tabular}{|c|c|c|c|c|c|c|c|c|c|c|}
			\hline
			No. &$M_\nu$ & $\eta^2 $ & $H_0$ & $\Omega_c+\Omega_b$ & $\Omega_\nu$ &$\Omega_\Lambda$ & $\sigma_8$ & $n_s$ & $A_s \,(10^{-9})$\\
			\hline
			\hline
			A1 & 0.06 & 0 & 67.37 & 0.3141 & 0.00140 & 0.6845 & 0.814 & 0.965 & 2.10 \\
			A2 & 0.06 & 0.253 & 68.13 & 0.3107 & 0.00142 & 0.68788 & 0.818 & 0.969 & 2.11\\
			A3 & 0.06 & 1.012 & 70.49  & 0.3009 & 0.00145 & 0.69765 & 0.833 & 0.982 & 2.15 \\
			\hline
			B1 & 0.15 & 0 & 66.43 & 0.3236 & 0.0036 & 0.6728 & 0.794 & 0.964 & 2.10 \\
			B2 & 0.15 & 0.253 & 67.12 & 0.3208 & 0.00365 & 0.67555 & 0.799 & 0.968 & 2.11\\
			B3 & 0.15 & 1.012 & 69.44 & 0.3106 & 0.00375 & 0.68565 & 0.811 & 0.981 & 2.15\\
			\hline
            C1 & 0.24 & 0 & 65.46 & 0.3337 & 0.00595 & 0.66035 & 0.775 & 0.963 & 2.11\\
			C2 & 0.24 & 0.253 & 66.17 & 0.3308 & 0.00600 & 0.6632 & 0.778 & 0.967 & 2.12\\
			C3 & 0.24 & 1.012 & 68.39 & 0.3192 & 0.00618 & 0.67462 & 0.790 & 0.981 & 2.15\\
			\hline
		\end{tabular}
		\caption{\label{tab:simparam} N-body simulation parameters.}
		\end{center}
\end{table}

Simulation snapshots are generated using our modified version of $\texttt{Gadget2}$ to incorporate the neutrino effects. We made 9 runs, each with $1024^3$ particles and a volume of over $(1000 h^{-1}\mathrm{Mpc})^3$ with a mass resolution of $8.15\times10^{10}h^{-1} M_\odot$. 
The initial conditions for the N-body simulations are generated using \texttt{MUSIC} with second-order Lagrangian corrections,
while the initial conditions of CDM and neutrino power spectra are obtained from the transfer function generated by \texttt{CAMB}, at the initial redshift $z=49$. 

To capture the halo assembly history, we stored 128 snapshots between $z=3$ to $z=0$ so that we can track potentially small changes of $a_{1/2}$ due to the neutrinos. The halo catalog is constructed using \texttt{Rockstar} \cite{rockstar}. The halo radius is defined to be the radius where the over-density equals $\Delta=200\rho_c$, where $\rho_c$ is the critical density of the universe. \texttt{Rockstar} is a 6D phase space friends-of-friends halo-finding algorithm, which specializes in identifying subhalos and tracking merger events. 
The halo merger tree is constructed by linking halos across different time steps using \texttt{Consistent-trees} \cite{ct} together with \texttt{Rockstar}. Finally we implement the calculation of $a_{1/2}$, $s$ and $\ell(X)$ with the built-in tool \texttt{read\_tree} inside \texttt{Consistent-trees}.

%% file: Maintext/halo-stat.tex
\section{Halo assembly statistics}
\label{sec:halo}

\subsection{Halo merger tree}
\begin{figure}[!h]
\begin{center}
\begin{subfigure}[h]{0.48\columnwidth}
        \includegraphics[width=\columnwidth]{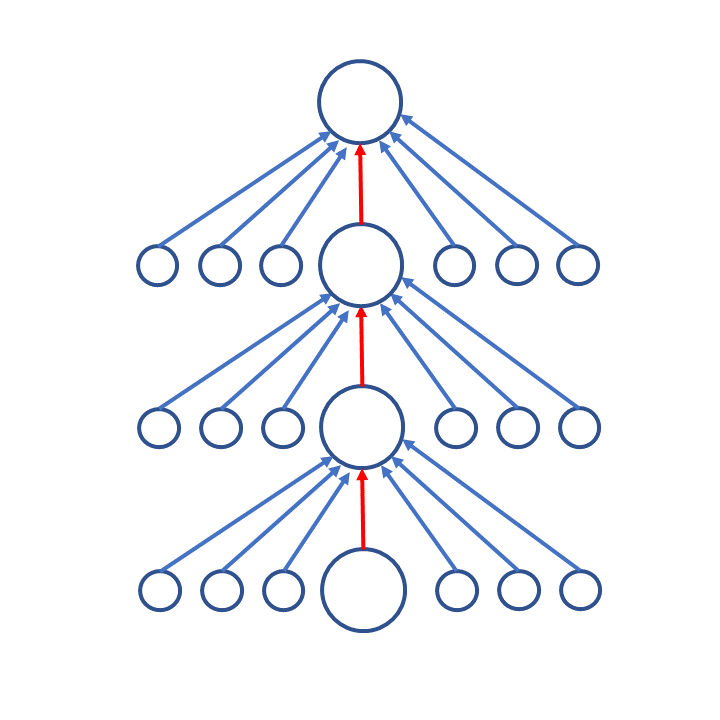}
        \caption{Halo assemble by smooth accretion}
\end{subfigure}
\begin{subfigure}[h]{0.48\columnwidth}
        \includegraphics[width=\columnwidth]{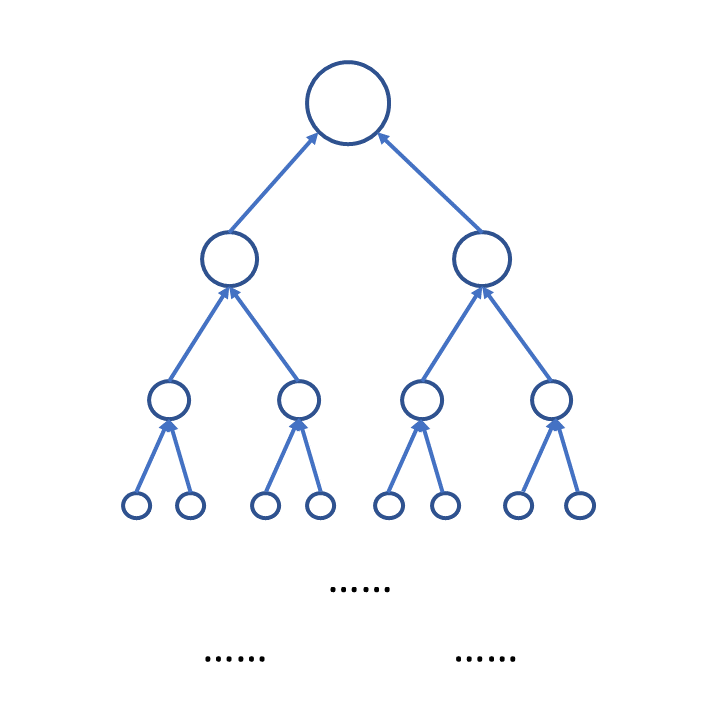}
        \caption{Halo assemble by major merger}
\end{subfigure}
	    \caption{\label{fig:MT} Illustration of two extreme types of halo merger trees \cite{tree entropy}.} 
\end{center}
\end{figure}

Dark matter halos can grow in two different ways: by accreting nearby matter or annexing other nearby self-bounded halos. A simple illustration of a halo merger tree is shown in Figure \ref{fig:MT}. For every halo existing at scale factor $a=1$ as the root, the halo merger tree branches out for each of its progenitors, reaching to the past and repeating until no progenitor is found. Once we construct the halo merger tree, the assembly history of a halo is specified, and the evolution of every halo property, such as halo mass, spin and concentration is captured.

Although the halo merger tree is a powerful tool to visualize how the halos assemble, it is not easy to compare two halo merger trees directly. Therefore, we use three parameters to quantify the characteristics of a halo merger tree:
the rate of growth of the halo mass, the fraction of the mass of a halo coming from mergers, and the number of protohalos merging into a single halo we observe today.
These characteristics of a halo assembly can be captured in the halo formation time $a_{1/2}$, tree entropy $s$, and halo leaf function $\ell(X)$ as we will see.

\subsection{Halo formation time $a_{1/2}$}
To characterize the mass growth rate of a halo, we can record its mass, trace one step back to the merger tree, pick the most massive progenitor (main progenitor) and repeat. It is a reduced representation of the halo assembly history called the mass accretion history (MAH) of the main branch.

We then define the halo formation time $a_{1/2}$ as the latest scale factor at which the main-branch halo reaches half of its current mass, i.e.,
\begin{equation}
    a_{1/2} = \mathrm{Max}(a') \mid M(a') = M(a=1)/2,
\end{equation}
where $M(a')$ is the main-branch halo mass at the scale factor $a'$.
We follow this traditional definition of halo formation time to quantify the rate of mass accretion, since $a_{1/2}$ was shown to be most correlated with the present-day Navarro–Frenk–White (NFW) concentration $c$ independent of the halo mass \cite{Kuan}.

Although we cannot observe the MAH of a particular halo in sky surveys, there is an observational proxy for $a_{1/2}$. 
It is known that the mass fraction of the main substructure $f_{main}$ is tightly correlated with $a_{1/2}$ in high-resolution N-body simulations \cite{Wang_2011_2}; 
the relationship is robust for different masses of the host halos. 
Using halo abundance matching (HAM), we can relate $f_{main}$ with $f_*$, which is the stellar mass fraction of the central galaxy.
The Sloan Digital Sky Survey (SDSS) data \cite{Wang_2011} shows a strong correlation between $f_*$ and galaxy properties such as color and star formation rate \cite{Lim_2015}. 
As a result, the neutrino effects on MAH, quantified by $a_{1/2}$, can be measured by direct observables.

Intuitively, as the neutrino masses suppress structure formation, halos should grow slower compared to those in the $\Lambda$CDM universe with zero neutrino mass, and we expect to see a delay in $a_{1/2}$ that depends on $M_\nu$. We do a simple quadratic fit on the middle panel of Figure 1 in \cite{Lim_2015} to get a relation between $a_{1/2}$ and $f_*$.
\begin{align}
\label{eq:emp_law}
     z_{1/2} = \frac{1}{a_{1/2}} -1 = (0.446\pm0.06) (\log_{10}(f_*))^2 + (2.4\pm0.3) \log_{10}(f_*) + (3.9\pm0.3)
\end{align}
Assuming a typical $a_{1/2}$ value of 0.5, Eq.(\ref{eq:emp_law}) implies a $1\%$ delay in $a_{1/2}$ compared to that in the $\Lambda$CDM universe with zero neutrino mass would result in a $1.4\%$ decrease in $\log_{10}(f_*)$.

\subsection{Tree entropy $s$}
Besides MAH, the merger history is another way to describe the halo assembly. Two halos may have similar MAH but entirely different merger histories (see Figure \ref{fig:ass_hist}). 

To quantify the merger history, the concept of tree entropy $s(a)$ is adopted \cite{tree entropy}. This dimensionless parameter captures the mass ratios of the mergers and the complexity of the merger tree geometry. Zero tree entropy $(s=0)$ corresponds to a tree with a single branch, i.e., with no merger. The maximum tree entropy $(s=1)$ corresponds to a fractal history of equal-mass binary mergers. The evolution of $s$ is calculated as follows: 

\begin{figure}[!ht]
	\begin{center}
	\includegraphics[width=0.8\columnwidth]{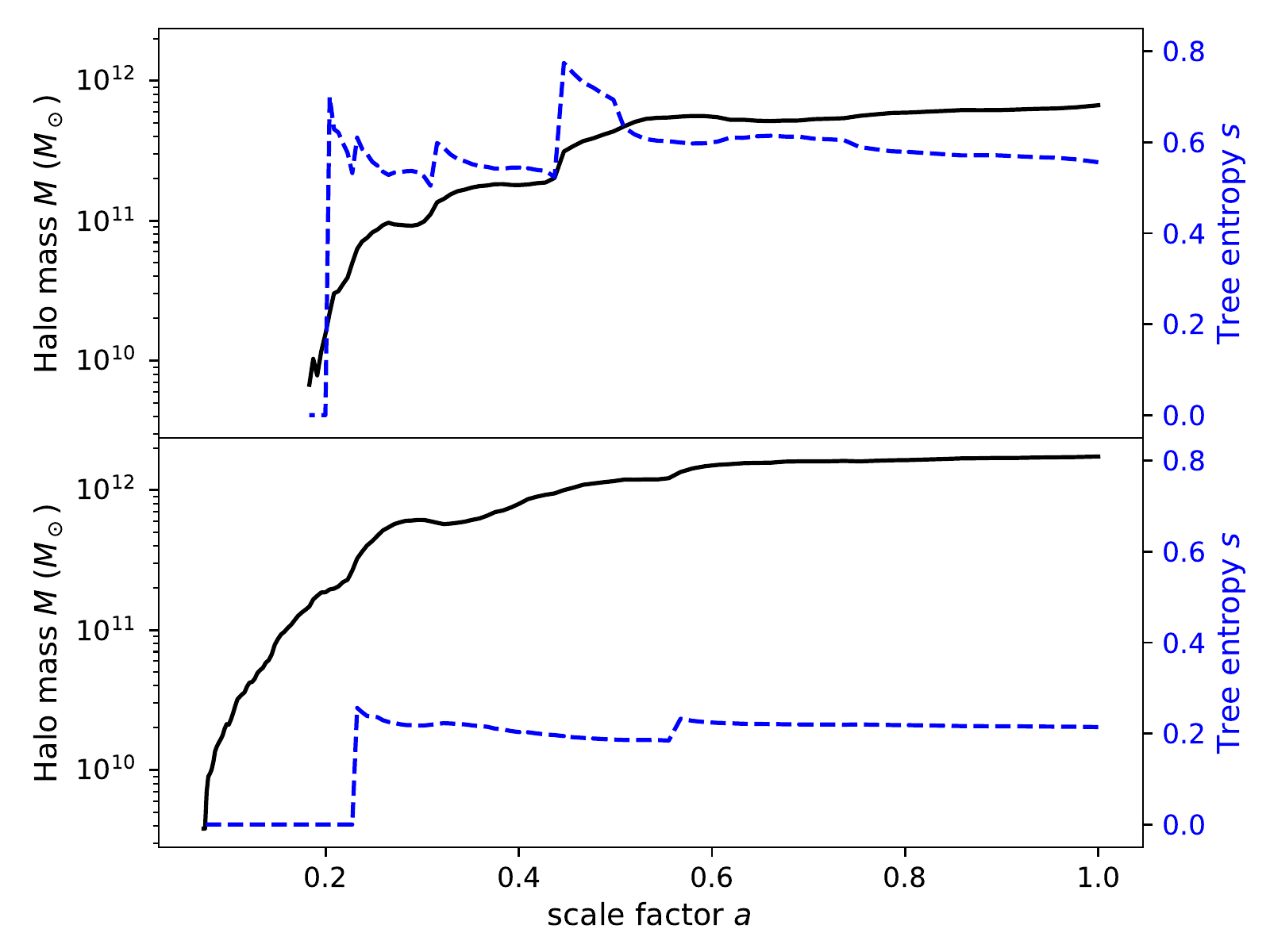}
	\caption{\label{fig:ass_hist}Assembly histories of two halos: the main-branch halo mass $M$ (black solid line) and tree entropy $s$ (blue bashed line) are plotted against the scale factor $a$. Both halos have similar formation time and final masses, but they have different merger histories. In the top panel, there is a tree entropy spike from $0$ to $\sim 0.75$ at $a=0.2$, indicating a binary merger with almost equal mass. The bottom panel shows a halo growing by smooth accretion instead. Two minor mergers occur, but they do not make up a significant mass fraction, and therefore the tree entropy remains small.}
	\end{center}
\end{figure}
\paragraph{Halos without progenitor} 
We set the initial tree entropy $s_{init} = 0$ for any halo without progenitor; these protohalos are formed by smooth accretion.

\paragraph{Mass growth by merger} For $n$ halos that merge together, each with mass $m_i$ and tree entropy $s_i$, the new tree entropy for the merged halo $s_{new}$ is calculated by:
\begin{subequations}
\label{S_equ}
\begin{align}
    x_i &= m_i/ \sum_{i=1}^n m_i ,\\
    H &= -f \sum_{i=1}^n x^\alpha_i \ln{x_i} , \\
    s_{new} &= H + (1+bH+cH^2) \sum_{i=1}^n x^2_i(s_i - H),
\end{align}
\end{subequations}
where $f=e(\alpha-1),\, b=(2-\gamma)/f,\, c= (1-\beta)e^{1/(\alpha-1)}-1-b$ are normalization constants. The 3 parameters $\alpha, \beta$ and $\gamma$ in turn govern the behavior of the tree entropy. For instance, $\alpha$ controls the impact of the merger order.
For a lower value of $\alpha$, a high-order merger ($n$ large) produces more tree entropy relative to a low-order one ($n$ small).
We would like binary mergers $(n=2)$ to have greater impact compared to triple mergers $(n=3)$, as the latter are more "accretion-like" than the former. This condition alone will fix $\alpha=1+ 1/ \ln(2)$.
$\beta$ on the other hand, controls the impact of the most destructive merger (equal-mass binary mergers) on $s$. $\gamma$ is related to the tree entropy loss when the halo is accreting mass smoothly.

\paragraph{Mass growth by accretion} 
The evolution of $s$ for smooth accretion of mass $\Delta m = M(t_2) - M(t_1)$, assuming no merger occurs between time $t_1$ and $t_2$, needs to be consistent with Eq.(\ref{S_equ}). We break down the smooth accretion as a series of consecutive $n^{th}$-order "mergers" $p$ times, with each "merging halo" having a mass $\delta m = \Delta m/[p(n-1)]$. In the limit of $p \to \infty$, regardless of $n$, Eq.(\ref{S_equ}) becomes:
\begin{equation}
    s_{new} = \left( \frac{m}{m+\Delta m} \right) ^\gamma.
\end{equation}

Here we follow the default choices for $(\alpha,\beta,\gamma)=(1+1/ \ln(2),\, 3/4,\, 1/3)$ specified in \cite{tree entropy}.
Using the tree entropy, we can identify the merger-rich halos and select them for specific studies.
As massive neutrinos suppress the small-scale correlation due to their free-streaming effect, one might expect that the merger histories of halos would be altered significantly, as they are very sensitive to small-scale correlation. 

\subsection{Halo leaf function $\ell(X)$}
\begin{figure}[htp]
	\begin{center}
	\includegraphics[width=\columnwidth]{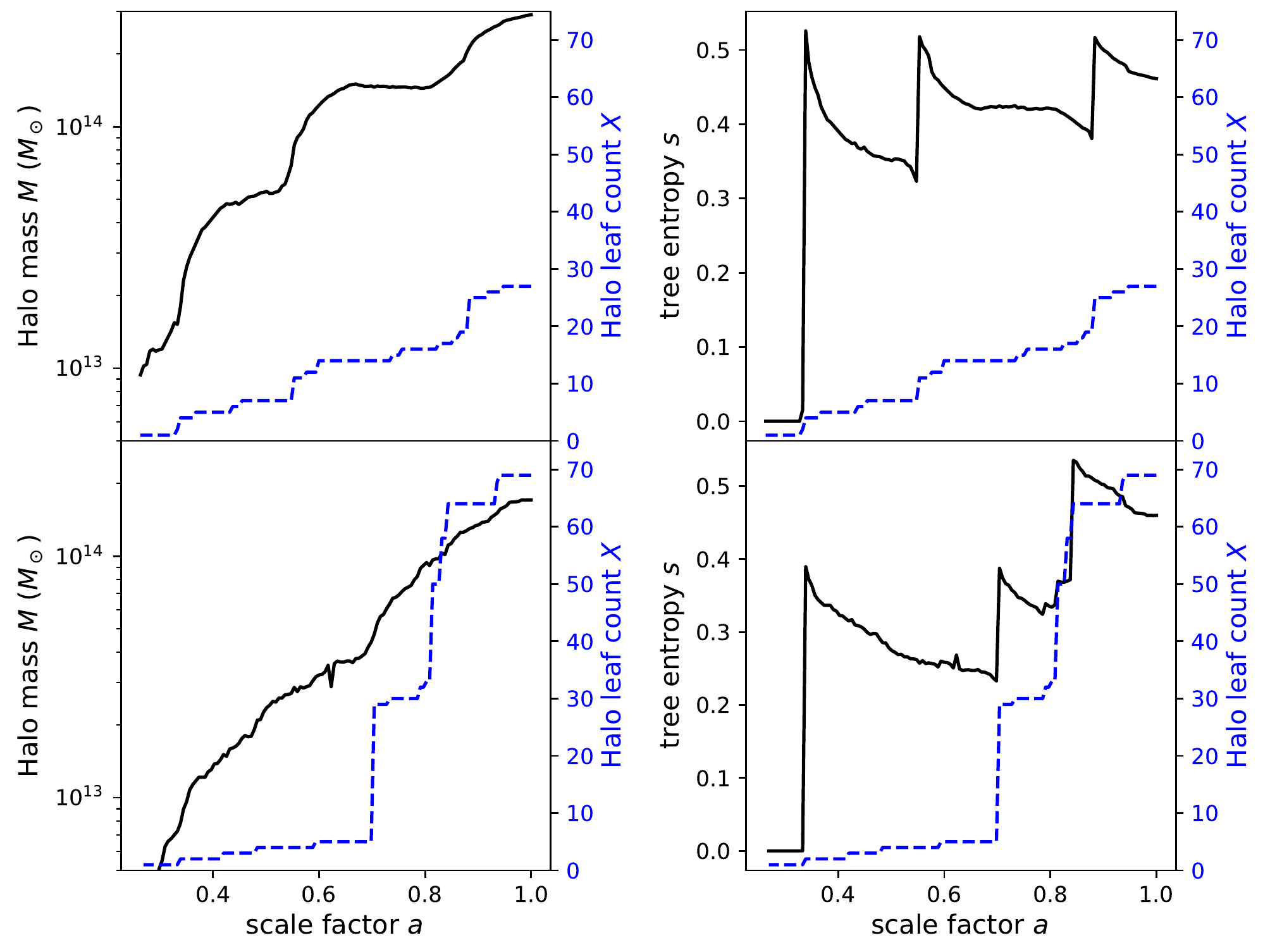}
	\caption{\label{fig:leaf_hist}Assembly histories of two halos: 
    The left panel is similar to Figure \ref{fig:ass_hist}.
	In the right panel,
	the tree entropy $s$ (black solid line) and halo leaf count $X$ (blue bashed line) are plotted against the scale factor $a$. 	
	Both halos have similar final masses and tree entropies $s\approx0.46$, but they have different $X$. 
	In the top panel, the halo gains its mass through three major mergers, shown by the spikes of $s$: each merger brings a huge impact to $s$ but only a small number of halo leaf counts to the main branch.
	The bottom panel shows a halo growing by many minor mergers instead. A huge spike of $X$ at $a=0.7$ is accompanied by only a small change in $s$, implying that the merger is minor, but the merged halo inherits the tree entropy from its entropy-rich progenitor.}

	\end{center}

\end{figure}

To quantify the merger histories of halos, we can also count the number of leaves (protohalos) in the halo merger trees. 
The halo leaf count $X$ is independent of $a_{1/2}$ and $s$. 
Imagine two halos with similar mass accretion histories, which give them similar $a_{1/2}$. 
One of them gains its mass from major mergers (binary mergers with comparable masses), and one of them gains its mass from many minor mergers. 
The latter must have more leaves than the former since the mass of each merging halo is smaller. 
The same argument can be applied to $s$: two halos might have similar $s$, but their $X$ can be drastically different (See Figure \ref{fig:leaf_hist}).

We define the halo leaf function $\ell(X)$ to be the number of halos with more than $X$ leaves.
The halo leaf function $\ell(X)$ is conceptually similar to the halo mass function, for which the halos are binned into different mass bins. 

%% file: Maintext/result.tex
\section{Simulation results}
\label{sec:result}
\subsection{Neutrino effects on mean formation time $\bar a$}
We study the distribution of the halo formation time $a_{1/2}$ for halos with mass $M$ between $10^{13} M_\odot$ and $3\times 10^{13} M_\odot$. 
The choice of the mass range ensures the halos contain more than 100 simulation particles and enough halo samples to reduce the statistical error of $\bar a$.
The mean formation time $\bar a$ for different $M_\nu$ and $\eta^2$ are listed in Table \ref{tab:formtime}, showing changes that are small but significant. The changes of the $a_{1/2}$ distributions can be represented by $\bar a$ of the halos in this mass range.

\begin{table}[!h]
		\begin{center}
		\begin{tabular}{|c|c|c|c|c|}
			\hline
			No. & $\bar a$ & Std. Err. & Sample size \\
			\hline \hline
			A1 & 0.58530 & 0.00024 & 316378  \\
			A2 & 0.58350 & 0.00024 & 315481  \\
			A3 & 0.57483 & 0.00024 & 310334  \\
			\hline
			B1 & 0.59319 & 0.00023 & 325006  \\
			B2 & 0.59038 & 0.00023 & 324867  \\
			B3 & 0.58331 & 0.00024  & 321874  \\
			\hline
			C1 & 0.60059 & 0.00023 & 338174  \\
			C2 & 0.59872 & 0.00023 & 337104  \\
			C3 & 0.59076 & 0.00023  & 334514  \\		
			\hline
		\end{tabular}
		\caption{\label{tab:formtime} $\bar a = \langle a_{1/2} \rangle$ for halos with mass $10^{13}M_\odot < M <3\times10^{13} M_\odot$} for simulations with different $M_\nu$ and $\eta^2$  (see Table \ref{tab:simparam}).
		\end{center}
\end{table}

\begin{figure}[!h]
		\begin{center}
		\includegraphics[width=0.7\columnwidth]{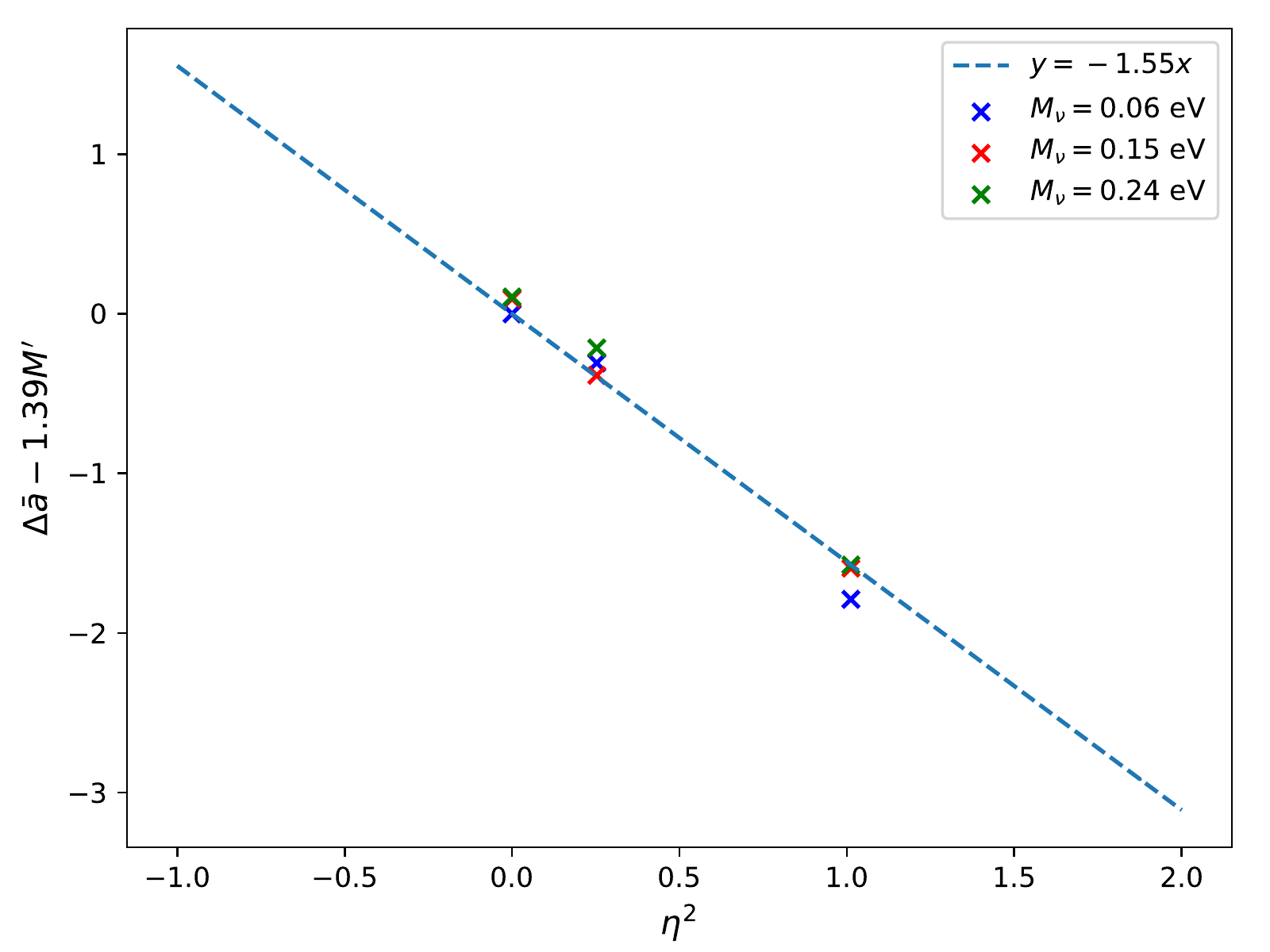}
		\caption{$\Delta \bar a - 1.39M'$ vs. $\eta^2$ (see Eq.(\ref{eq:reg_da})) for different $M_\nu$ in Table \ref{tab:formtime}. This combination of $\Delta \bar a$ and $M'$ should depend only on $\eta^2$ in the regression model, and the simple linear fit describes the simulation data quite well.}
		\label{fig:da_m}
		\end{center}
\end{figure}

To parameterize the effects of neutrinos on $\bar a$, we fit the fractional change of $\bar a$ as a function of $M_\nu$ and $\eta^2$. Here we choose $M_\nu = 0.06$ eV and $\eta^2=0$ as the baseline for comparison. We define:
\begin{equation}
    \Delta \bar a(M_\nu,\eta^2) \equiv \frac{\bar a(M_\nu, \eta^2) - \bar a(0.06\mathrm{ \,eV}, 0)}{\bar a(0.06\mathrm{\,eV}, 0)}.
\end{equation}
The regression model is:
\begin{align}
\label{eq:reg_da}
    \Delta \bar a(M_\nu,\eta^2) [\%] &= C_m \left(\frac{M_\nu-0.06\mathrm{\, eV}}{0.1\mathrm{\, eV}}\right) + C_\eta\eta^2, \nonumber \\
    \Delta \bar a(M',\eta^2) [\%]&= (1.39\pm0.06)M' -(1.55\pm0.07)\eta^2,
\end{align}
where $M'=(M_\nu[\mathrm{0.1eV}]-0.6)$, the deviation of $M_\nu$ in units of $0.1\,\mathrm{eV}$ from our chosen baseline. We plot $\Delta \bar{a} - 1.39 M'$ in Figure \ref{fig:da_m}, which should depend on $\eta^2$.

\begin{figure}[!h]
		\begin{center}
		\includegraphics[width=0.7\columnwidth]{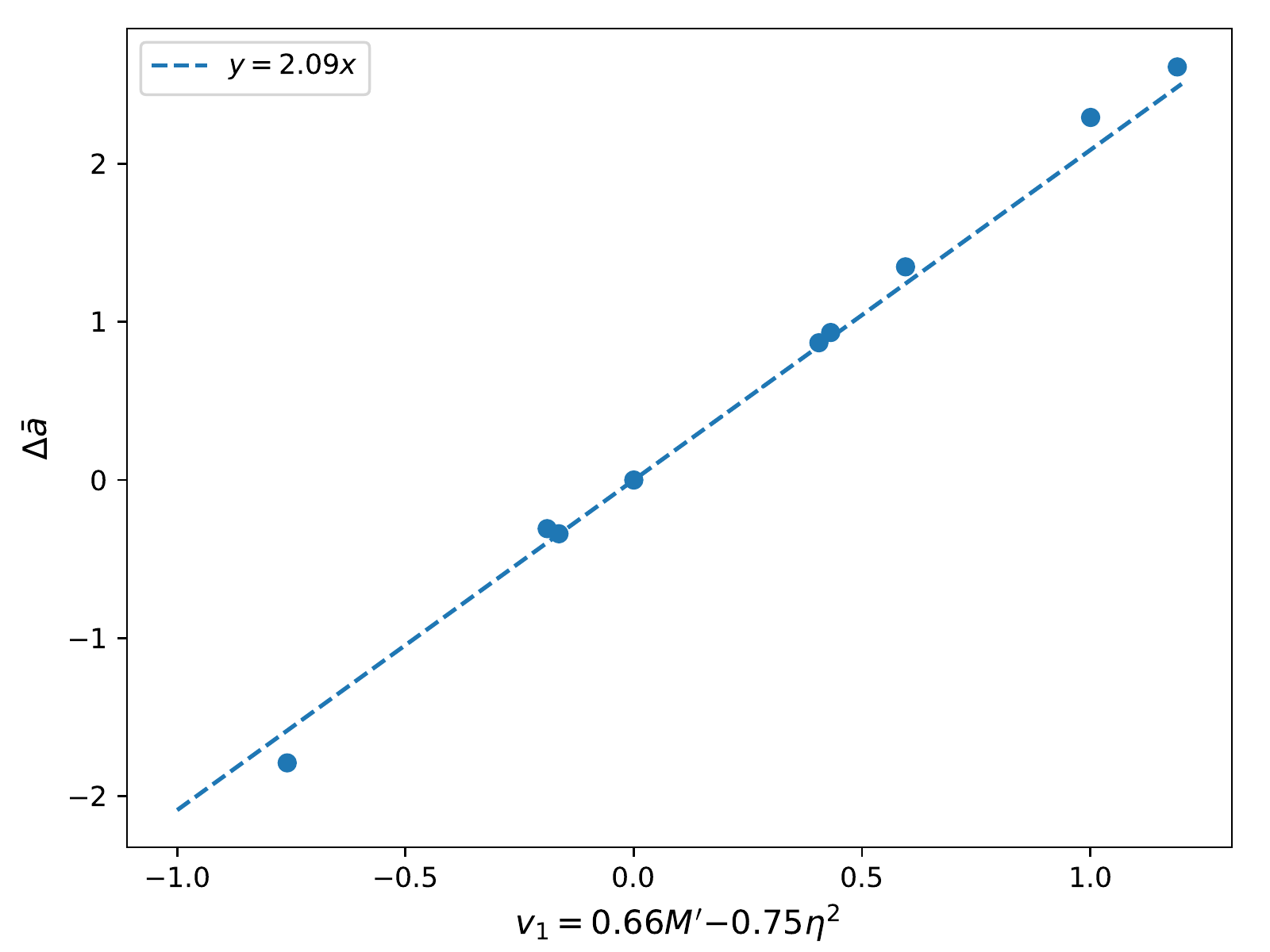}
		\caption{$\Delta {\bar a}$ vs. $v_1$ (see Eq. (\ref{eq:reg_v1})) for various $M_\nu$ and $\eta^2$, showing the quality of the regression model.}
		\label{fig:da_v}
		\end{center}
\end{figure}

Notice that we use $\eta^2$, not $\eta$ for the dependence of $\Delta \bar a$. 
This is because the regression of cosmological parameters suggest that the coefficient of $\eta^1$ is 0 within uncertainty (see Appendix \ref{app:posterior}).
Physically, the changes in the neutrino energy density are proportional to $\eta^2$ to the lowest order when neutrino asymmetries are introduced.
Finite $M_\nu$ and $\eta^2$ have opposite effects on $\Delta \bar a$. A previous study also showed similar results on the cosmological parameters and matter power spectrum \cite{Carton}.

To unveil the possible correlation between $M'$ and $\eta^2$ in the fitting of $\Delta \bar a$, we compute the covariance matrix of the fitting, and the off-diagonal elements are significant indeed. We define $v_1, v_2$ as:
\begin{equation}
\label{eq:rot}
    \begin{cases}
    v_1 = M'\cos \theta  - \eta^2 \sin \theta,  \\ 
    v_2 = M' \sin \theta  + \eta^2\cos \theta.
    \end{cases}
\end{equation}
We find that $\theta = 0.27\pi$ will minimize the off-diagonal terms of the covariance matrix in the new basis. The contribution of $v_1 = (0.66 M' - 0.75\eta^2)$ to $\Delta \bar a$ dominates over that of $v_2$.
\begin{equation}
\label{eq:reg_v1}
    \Delta \bar a(v_1, v_2) [\%] = (2.09\pm0.09) v_1 + (0.02\pm0.03) v_2.
\end{equation}
We plot $\Delta \bar a(v_1)$ in Figure \ref{fig:da_v}. The lack of its dependency on $v_2$ implies that we can only determine the combination of $M_\nu$ and $\eta^2$ as $v_1$ using $\Delta \bar a$. We hope to break this degeneracy between $M_\nu$ and $\eta^2$ using the merger histories of the halos.

\subsection{Neutrino effects on the mean tree entropy $\bar s$}
The tree entropy $s$ is also calculated for the halo catalog,
including all halos with mass $M>4\times10^{13} M_\odot$ and excluding those that never experienced any merger (i.e. $s=0$). Those halos would have more than 400 simulation particles, and thus the merger histories of halos can be captured accurately.
However, the changes in the mean tree entropy $\bar s$ for different $M_\nu$ and $\eta^2$, $\Delta {\bar s} (M', \eta^2) \equiv \bar s(M', \eta^2) / \bar s (0, 0)-1$, are barely significant. 
The shifts in the distribution of $s$ at a higher redshift (see Figure \ref{fig:s_dis}) are still fairly small.
The regression of $\Delta \bar s$ on $M'$ and $\eta^2$ gives:
\begin{equation}
    \Delta \bar s(M',\eta^2) [\%] = (-0.02\pm0.03) M' + (0.15\pm0.07) \eta^2.
\end{equation}

\begin{minipage}{\columnwidth}
\centering
  \begin{minipage}[b]{0.49\columnwidth}
    \centering
    \includegraphics[width=\columnwidth]{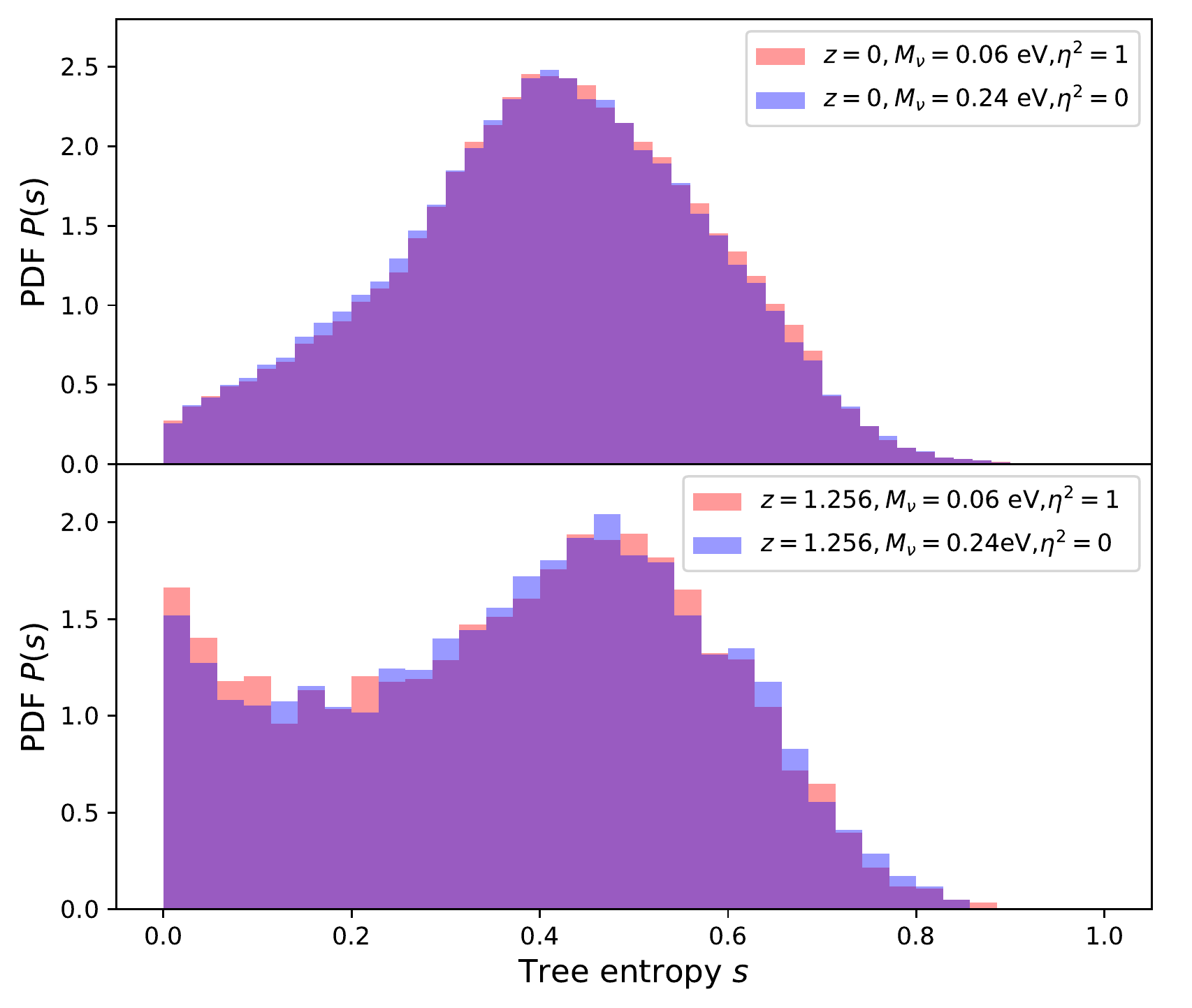}
    \captionof{figure}{\label{fig:s_dis}PDF of tree entropy for A3 and C1 at $z=0$ (upper panel) and 1.256 (lower panel).}
  \end{minipage}
  \hfill
  \begin{minipage}[b]{0.49\columnwidth}
    \centering
		\begin{tabular}{|c|c|c|c|c|}
			\hline 
			No. & $\bar s$ & Std. Err. & Sample size \\
			\hline \hline
			A1 & 0.40912 & 0.00054 & 94488  \\
			A2 & 0.40922 & 0.00054 & 94272  \\
			A3 & 0.41090 & 0.00054 & 93997  \\
			\hline
			B1 & 0.40773 & 0.00054 & 94421  \\
			B2 & 0.40921 & 0.00054 & 94566  \\
			B3 & 0.41135 & 0.00054 & 94101  \\
			\hline
			C1 & 0.40718 & 0.00054 & 94653  \\
			C2 & 0.40785 & 0.00054 & 94256 \\
			C3 & 0.41044 & 0.00054 & 93886  \\		
			\hline
		\end{tabular}
      \captionof{table}{ \label{tab:entropy_dis} Mean tree entropies for different simulations $\bar s = \langle s \rangle$ with their standard errors and sample sizes.}
    \end{minipage}
\end{minipage}
\vspace{0.5cm}

At $z=0$, all 9 runs show $\bar s\sim 0.4$ with overlapping error bars (See Table $\ref{tab:entropy_dis}$).
To reveal the neutrinos' effects on the merger histories of halos, we calculate $\bar s$ for different halo leaf counts $X$.

\begin{table}[!hbt]
		\begin{center}
\begin{tabular}{|c|c|c|c|c|c|}
			\hline 
			No & $X$ bin & $\bar s$ & Sample size \\
			\hline \hline
			A1 & $1<X\le6$ & 0.3216 & 16890  \\
			B1& $6<X\le30$ & 0.4208 & 68126  \\
			C1 & $30<X\le120$ & 0.4706 & 8425  \\
			\hline
			A1 & $1<X\le6$ & 0.3264 & 19079  \\
			B1 & $6<X\le30$ & 0.4211 & 66716  \\
			C1 & $30<X\le120$ & 0.4720 & 7758  \\
			\hline
			A1 & $1<X\le6$ & 0.3293 & 21226  \\
			B1 & $6<X\le30$ & 0.4226 & 65759  \\
			C1 & $30<X\le120$ & 0.4767 & 6926  \\
			\hline
		\end{tabular}
		\caption{$\bar s$ for different neutrino mass $M_\nu$ (A1: $M_\nu = 0.06$ eV, B1: $M_\nu = 0.15$ eV, C1: $M_\nu = 0.24$ eV. All with fixed $\eta^2 = 0$) with $X$ binning.}
		\end{center}
\end{table}

With the $X$ filter, we can see two general trends. Firstly, $X$ has a positive correlation with $\bar s$. This is intuitive as a larger $X$ would imply more merging events for the halo, and such a halo would more likely experience major mergers. Secondly, in each $X$ bin, $\bar s$ increases as $M_\nu$ increases, especially for halos with lower $X$. The reason that we do not see an overall $\bar s$ dependence on $M_\nu$ is that while a larger $M_\nu$ increases $\bar s$ for each $X$ bin, it also decreases the number of halos with large $X$, which have larger $s$, and the two effects compensate for each other.

Qualitatively, the increase in $\bar s$ with increasing $M_\nu$ for halos with a given $X$ can be understood as a result of the shorter neutrino free-streaming length, which leads to less clustering of halos. On average, the halo environment would be less dense and there would be less smooth accretion which would decrease $s$.

\subsection{Neutrino effects on the halo leaf function $\ell(X)$}
\begin{figure}[!h]
		\begin{center}
		\includegraphics[width=\columnwidth]{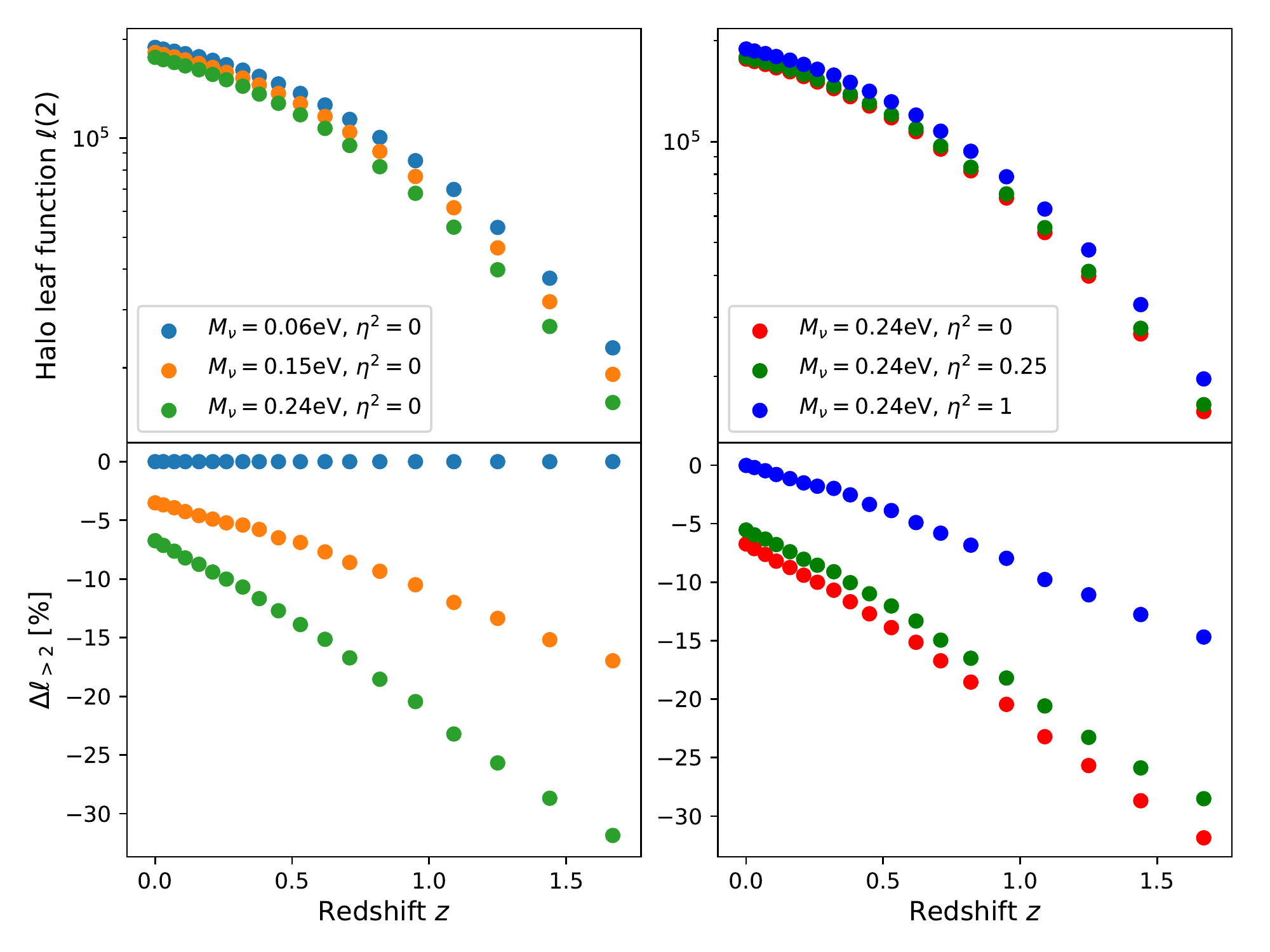}
		\caption{\label{fig:N_x}Number of halos containing more than 2 leaves ($\ell(2)$, top panel) and $\Delta \ell_{>2}$ (bottom panel) vs. $z$, for different $M_\nu$ and $\eta^2$.  As $M_\nu$ increases (left panel), the suppression of $\ell(2)$ at large $z$ is more severe, implying larger delay of  merger events. The effect of $\eta^2$ partially cancels with that of $M_\nu$ (right panel). }
		\end{center}
\end{figure}

The halo leaf function $\ell(X)$ for different $M_\nu$ and $\eta^2$ is calculated.
We define the fractional change of $\ell(X)$ as $\Delta \ell_{>X} \equiv \ell(M', \eta^2, X) /\ell(0, 0, X) - 1$.
We plot $\Delta \ell_{>2}$ vs. $z$ for various $M_\nu$ and $\eta^2$ in Figure \ref{fig:N_x}. We then do the linear fitting of $\Delta \ell_{>2}$ for different $z$:
\begin{equation}
\label{eq:reg_dN}
    \Delta \ell_{>2}(M', \eta^2)[\%] = D_m  M' + D_\eta\eta^2,
\end{equation}
and we rotate the basis $(M', \eta^2)$ to $(w_1, w_2)$ using $w_1 = M' \cos \phi\, -\, \eta^2 \sin \phi$ and $w_2 = M' \sin \phi \,+\, \eta^2 \cos \phi$ to minimize the off-diagonal terms in the covariance matrix:\
\begin{equation}
\label{eq:reg_dN_w}
    \Delta \ell_{>2}(w_1, w_2)[\%]= D_1 w_1 + D_2 w_2.
\end{equation}
\begin{table}[!h]
		\begin{center}
		\begin{tabular}{|c|c c|c|c c|}
			\hline
			$z$ & $D_m$ & $D_\eta$ & $\phi (\pi)$ & $D_1$ & $D_2$\\
			\hline \hline
			0.00 & $-3.44 \pm 0.21$ & $6.09 \pm 0.38$ & 0.337 & $-6.99 \pm 0.44$ & $-0.08 \pm 0.00$ \\
			0.03 & $-3.65 \pm 0.22$ & $6.30 \pm 0.39$ & 0.337 & $-7.28 \pm 0.44$ & $-0.17 \pm 0.02$ \\
			0.07 & $-3.87 \pm 0.25$ & $6.40 \pm 0.44$ & 0.338 & $-7.47 \pm 0.50$ & $-0.31 \pm 0.04$ \\
			0.11 & $-4.07 \pm 0.30$ & $6.43 \pm 0.53$ & 0.339 & $-7.59 \pm 0.60$ & $-0.47 \pm 0.07$ \\
			0.16 & $-4.26 \pm 0.36$ & $6.39 \pm 0.62$ & 0.337 & $-7.65 \pm 0.70$ & $-0.65 \pm 0.12$ \\
			0.21 & $-4.64 \pm 0.36$ & $6.70 \pm 0.60$ & 0.336 & $-8.11 \pm 0.68$ & $-0.84 \pm 0.14$ \\
			0.26 & $-5.12 \pm 0.32$ & $7.27 \pm 0.52$ & 0.334 & $-8.86 \pm 0.60$ & $-0.71 \pm 0.13$ \\
			0.32 & $-5.59 \pm 0.27$ & $7.94 \pm 0.44$ & 0.334 & $-9.68 \pm 0.51$ & $-0.77 \pm 0.11$ \\
			0.38 & $-6.15 \pm 0.29$ & $8.41 \pm 0.45$ & 0.330 & $-10.37 \pm 0.51$ & $-1.01 \pm 0.13$ \\
			0.45 & $-6.95 \pm 0.27$ & $9.11 \pm 0.40$ & 0.321 & $-11.42 \pm 0.46$ & $-0.99 \pm 0.14$ \\
			0.53 & $-7.51 \pm 0.31$ & $9.56 \pm 0.44$ & 0.313 & $-12.13 \pm 0.52$ & $-0.83 \pm 0.16$ \\
			0.62 & $-8.21 \pm 0.42$ & $9.93 \pm 0.53$ & 0.296 & $-12.86 \pm 0.64$ & $-0.80 \pm 0.21$ \\
			0.71 & $-9.06 \pm 0.46$ & $10.47 \pm 0.53$ & 0.277 & $-13.84 \pm 0.67$ & $-0.30 \pm 0.20$ \\
			0.82 & $-9.94 \pm 0.57$ & $11.23 \pm 0.62$ & 0.269 & $-14.99 \pm 0.81$ & $-0.03 \pm 0.24$ \\
			0.95 & $-11.27 \pm 0.50$ & $12.51 \pm 0.53$ & 0.262 & $-16.83 \pm 0.71$ & $0.35 \pm 0.20$ \\
			1.09 & $-12.20 \pm 0.84$ & $12.72 \pm 0.81$ & 0.244 & $-17.60 \pm 1.14$ & $0.92 \pm 0.25$ \\
			1.25 & $-13.43 \pm 1.05$ & $13.83 \pm 0.99$ & 0.241 & $-19.25 \pm 1.41$ & $0.89 \pm 0.29$ \\
			1.44 & $-15.89 \pm 0.80$ & $16.00 \pm 0.74$ & 0.239 & $-22.54 \pm 1.07$ & $0.78 \pm 0.17$ \\
			1.67 & $-18.63 \pm 0.65 $ & $ 18.36 \pm 0.60$ & 0.236 & $ -26.15 \pm 0.87 $ & $ 0.63 \pm 0.11$\\
			\hline
		\end{tabular}
		\caption{\label{tab:dN_reg} Regression result for $\Delta \ell_{>2}$ at different $z$ (see Eqs.(\ref{eq:reg_dN}) and (\ref{eq:reg_dN_w})).}
		\end{center}
\end{table}
\begin{figure}[!h]
		\begin{center}
		\includegraphics[width=0.71\columnwidth]{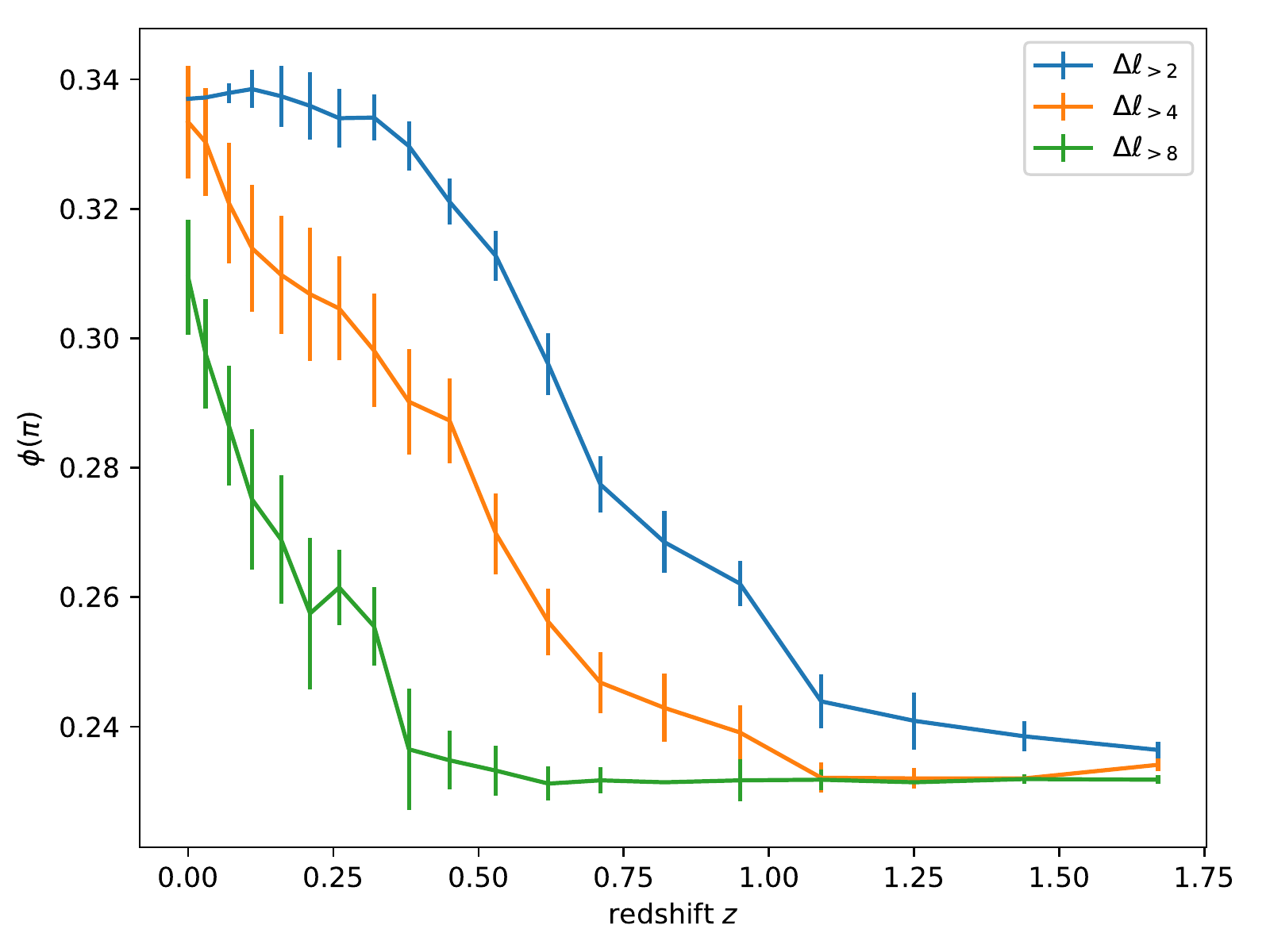}
		\caption{\label{fig:phi} Values of $\phi$ in Eq.(\ref{eq:reg_dN_w}) for $\Delta \ell_{>2}$,  $\Delta \ell_{>4}$ and $\Delta \ell_{>8}$. All of them go through a similar transition, though at different redshifts $z$. 
		To obtain the error bars for $\phi$, we fit $\Delta \ell_{>X}$ as a function of $\phi$ using Eq.(\ref{eq:reg_dN_w}) with $D_1, D_2$ fixed at their best-fit values in Table \ref{tab:dN_reg}.}
		\end{center}
\end{figure}

We first focus on the regression model Eq.(\ref{eq:reg_dN}). The signs of $D_m$ and $D_\eta$ in Table \ref{tab:dN_reg} are opposite, implying that the effects of $M_\nu$ and $\eta^2$ on $\ell(2)$ are opposite. 
$M_\nu$ suppresses $\ell(2)$, while $\eta^2$ enhances it. 
This may be due to the modified Hubble expansion. 
In Appendix \ref{app:posterior}, we show that a larger $M_\nu$ ($\eta^2$) leads to a smaller (larger) $H_0$.
A faster expansion leads to fewer halo mergers,
resulting in a universe with more halos with two or less leaves.

If we rotate the basis to $(w_1, w_2)$, the contribution of $w_2$ is again minimal, and the rotation angle $\phi$ is redshift-dependent. $\phi$ stays at $0.24\pi$ for high redshift and gradually increases to $0.34\pi$ before it stabilizes again at $z\sim0.2$. 
Like $\Delta \bar a$, $\Delta \ell_{>2}$ can only be used to determine a specific combination of $M_\nu$ and $\eta^2$. However, it provides different combinations at different redshifts. Therefore, the values of $\Delta \ell_{>2}$ at different redshifts can constrain both $M_\nu$ and $\eta^2$.
Since the contribution of $w_2$ is small, $\Delta \ell_{>2}(z)\approx D_1(z) w_1[\phi(z)]$. $D_1(z)$ controls the overall suppression compared to the baseline, while $\phi(z)$ controls the relative weights of $M_\nu$ and $\eta^2$.

The transition of $\phi$ is due to the different decay rates of $D_m$ and $D_\eta$. 
Between $z=1.67$ and $z=0$, $D_m$ drops by a factor of $5$, whereas $D_\eta$ only drops by a factor of $3$. 
However, the value of $\phi$ changes quite rapidly at $z\sim0.7$, and there are no correspondingly sudden changes of $D_m$ and $D_\eta$. A deeper understanding of the transition in $\phi$ is an interesting future work. 

We repeated the calculation for $\ell(4)$ and $\ell(8)$, and similar transitions in $\phi$ are observed (see Figure \ref{fig:phi}) even though they are delayed compared to that for $\ell(2)$.

%% file: Maintext/conclusion.tex
\section{Conclusion}
\label{sec:conclusion}
In this paper, we study the effects of neutrino masses and asymmetries on the halo assembly, including mass accretion and merger histories. Our simulations include effects of not only the neutrino free-streaming but also the refitted cosmological parameters with finite $M_\nu$ and $\eta^2$ from the Planck 2018 data.

Our simulations show that the neutrino asymmetry parameter $\eta^2$ and the sum of neutrino masses $M_\nu$ both have noticeable effects on the mean halo formation time $\bar a$ and halo leaf function $\ell(X)$.
While a larger $M_\nu$ would delay $\bar a$ and suppress $\ell(2)$, a non-zero $\eta^2$ has the opposite effect.
The mean tree entropy $\bar s$, on the other hand, is not sensitive to $M_\nu$ and has only a weak dependence on $\eta^2$. Further investigation is needed to separate the neutrino effects on the merger order and merger mass ratio.

We also present a linear regression of deviations of $\bar a$ and  $\ell(2)$ from their baseline on $M_\nu$ and $\eta^2$. Rotating to the uncorrelated bases ($\{v_1, v_2\}$ and $\{w_1, w_2\}$), we find that both $\Delta \bar a$ and $\Delta \ell_{>2}$ only depend on one combination of $M_\nu$ and $\eta^2$, $v_1$ and $w_1$, respectively. Therefore, we cannot constrain both $M_\nu$ and $\eta^2$ using only $\bar {a}$ or $\ell(2)$.

However, if we follow $\Delta \ell_{>2}$ as a function of $z$, we find that $w_1$ depends on $z$, 
and the rotation angle $\phi$ specifying the relative contributions of $M_\nu$ and $\eta^2$ experiences a smooth transition from $0.24\pi$ to $0.34\pi$. 
Similar but delayed $\phi$ transitions occur for $\Delta \ell_{>4}$ and $\Delta \ell_{>8}$ as well.
Further investigation is needed to determine the cause of such a transition.

There are some proxies to measure the $a_{1/2}$ of a halo in sky surveys. A previous study using N-body simulations with a semi-analytical model has established an empirical relation of the stellar mass ratio in the central galaxy $f_*$ and $a_{1/2}$ of its host halo \cite{Lim_2015}. 
However, the law assumes a fixed cosmology and some baryonic physics. We still need to study how the relation changes in different cosmologies, and how those baryonic physics interact with the neutrino physics even though they govern the LSS in different length scales.

The halo leaf count $X$ is a new construct to measure the number of merger events to form a single halo, and the halo leaf function $\ell(X)$ quantifies how many such merger-rich halos exist. Conceptually, the halo leaf count $X$ could be inferred by other halo observables, such as halo spin. One would expect a halo with a very large spin to have a large $X$ as it is likely formed by merging with other halos. We can then correlate $\ell$ with $X$ in the halo catalog. This is an interesting follow-up project.

When studying neutrinos using cosmological probes, the matter power spectrum is often used \cite{Carton, Yvvone, Yvvone fluid, nuconcept} as it is easy to compute and the effects are significant even if we only consider the neutrino free-streaming effect. 
However, as shown in \cite{halo bias}, the halo observables, such as the halo spin and NFW concentration often show unnoticeable differences for a wide range of $M_\nu$. 
The halo assembly history may be a good starting point to explore the effects of neutrinos on halo properties.
Here, we have shown that $a_{1/2}$ and $\ell_{>2}$ can be used to break the parameter degeneracy between $M_\nu$ and $\eta^2$ so that they can both be measured in principle.

%% file: Maintext/linear-evo.tex
\section{Neutrino over-density linear evolution }
\label{app:phi}
We follow the derivation in \cite{Carton, LBE},  starting with the Vlasov equation:
\begin{equation}
\label{eq:a1}
    \frac{dF_\nu}{dt} = \pdv{F_\nu}{t} + \pdv{F_\nu}{r_i} \frac{dr_i}{dt} + \pdv{F_\nu}{p_i} \frac{dp_i}{dt} = 0,
\end{equation}
where $F_\nu(r_i,p_i,t)$ is the neutrino distribution function. Since we are considering a linear evolution equation, $F_\nu$ can be separated into the unperturbed Fermi-Dirac term $f_\nu^0(p)$ and a first-order perturbation $f'_\nu(\bold r, \bold p)$, i.e., $F_\nu = f_\nu^0 + f'_\nu$. In the following derivation, we will keep $f'_\nu$ up to the first order. In \texttt{Gadget2}, the Newtonian potential used is non-relativistic. Therefore,
\begin{equation}
\label{eq:a2}
    \bold{\dot p} = m \nabla \phi = -Gm \int \rho_t \frac{\bold{r - r'}}{|\bold{r-r'}|^3 } d^3r',
\end{equation}
where $\rho_t$ is the total matter energy density, including neutrinos and CDM. Substituting Eq.(\ref{eq:a2}) into Eq.(\ref{eq:a1}) and transforming to the co-moving coordinates with the follow rules:
\begin{align}
    d\chi &\equiv \frac{dt}{a^2(t)} \nonumber, \\
    \bold x &\equiv \frac{\bold r}{a(t)}, \\
    \bold u &\equiv \frac{d \bold x}{d\chi} = a(t) \bold v - \dot a(t) \bold r \nonumber,
\end{align}
we arrive at:
\begin{equation}
\label{eq:a4}
    \frac{1}{a^2}\pdv{f'_\nu}{\chi} + \frac{\bold u}{a^2} \cdot \pdv{f'_\nu}{\bold x} - \ddot a a \bold x \cdot \pdv{f_\nu^0}{\bold u} - Ga^2 \pdv{f_\nu^0}{\bold u} \cdot \int \rho_t \frac{\bold{x - x'}}{|\bold{x-x'}|^3 } d^3x' = 0.
\end{equation}
Recognizing the Dirac delta function in the integrand in Eq.(\ref{eq:a4}), we can combine it with the third term as it contains $\bold x$, and we can eliminate $\ddot a$ using the Friedmann equation:
\begin{align}
\label{eq:a5}
    \frac{\ddot a}{a} &= - \frac{4\pi G}{3} \bar \rho_t \nonumber,\\
    \frac{4\pi}{3} \bold x &= \int \frac{\bold{x - x'}}{|\bold{x-x'}|^3 } d^3x' \nonumber ,\\
    \ddot a \bold x &= -Ga \bar \rho_t \int \frac{\bold{x - x'}}{|\bold{x-x'}|^3 } d^3x',
\end{align}
where $\bar \rho_t$ is the mean total matter density. We then put Eq.(\ref{eq:a4}) to Eq.(\ref{eq:a5}) and multiply both sides by $a^2$ to obtain
\begin{equation}
\label{eq:a6}
    \pdv{f'_\nu}{\chi} + \bold u \cdot \pdv{f'_\nu}{\bold x} - Ga^4 \pdv{f_\nu^0}{\bold u} \cdot \int \bar \rho_t \delta_t( \chi, \bold x') \frac{\bold{x - x'}}{|\bold{x-x'}|^3 } d^3x' = 0,
\end{equation}
since by definition $\bar \rho_t \delta_t = \rho_t - \bar \rho_t$. Next we apply Fourier transform to Eq.(\ref{eq:a6}),  denoting $\Tilde{f}(\chi, \bold k ,\bold u) = \mathcal{F}[f(\chi, \bold x, \bold u)]$,
\begin{equation}
\label{eq:a7}
    \pdv{\Tilde{f'_\nu}}{\chi} + i\bold k \cdot \bold u \Tilde{f'_\nu} - Ga^4 \pdv{f^0_\nu}{\bold u} \int d^3x' \, \rho_t \delta_t(\chi, \bold x') \int d^3x\, e^{-i\bold k \cdot \bold x} \frac{\bold{x - x'}}{|\bold{x-x'}|^3 }  = 0.
\end{equation}
The last integral in Eq.(\ref{eq:a7}) can be evaluated,
\begin{equation}
    \int e^{-i\bold k \cdot \bold x} \frac{\bold{x - x'}}{|\bold{x-x'}|^3 } d^3x = -4\pi i \frac{\bold k}{k^2} e^{-i\bold k \cdot \bold x'}.
\end{equation}
Therefore, we have
\begin{equation}
    \pdv{\Tilde{f'_\nu}}{\chi} + i\bold k \cdot \bold u \Tilde{f'_\nu} + 4 \pi i Ga^4 \frac{\bold k}{k^2} \cdot \pdv{f^0_\nu}{\bold u} \bar \rho_t \Tilde{\delta_t}(\chi, \bold k) = 0.
\end{equation}
We then multiply both sides by $ e^{i \bold k \cdot \bold u \chi} $ and group the first two terms as a total derivative before integrating over co-moving time $\chi$,
\begin{align}
\label{eq:a10}
    \pdv{}{\chi}[\Tilde{f'_\nu} e^{i\bold k \cdot \bold u \chi}] + 4 \pi i Ga^4 e^{i \bold k \cdot \bold u \chi} \frac{\bold k}{k^2} \cdot \pdv{f^0_\nu}{\bold u} \bar \rho_t \Tilde{\delta_t}(\chi, \bold k) &= 0  \nonumber, \\
    \Tilde{f'_\nu}(\chi, \bold k, \bold u) + \int^\chi_0 4\pi iGa^4 e^{-i \bold k \cdot \bold u (\chi-\chi')} \frac{\bold k}{k^2} \cdot \pdv{f^0_\nu}{\bold u} \bar \rho_t \Tilde{\delta_t}(\chi, \bold k) d\chi' &= \Tilde{f'_\nu}(0, \bold k, \bold u) e^{-i \bold k \cdot \bold u \chi}.
\end{align}
Now Eq.(\ref{eq:a10}) is recognizable as Eq.(\ref{eq:LBE}). With an initial perturbation and the total over-density $\bar \rho_t \delta_t \equiv \bar \rho_{cdm} \delta_{cdm} + \bar \rho_\nu \delta_\nu$, we can evolve the neutrino perturbation function. Now we convert the distribution function $f'_\nu$ to over-density $\delta_\nu$ by integrating over the momentum space:
\begin{equation}
\label{eq:a11}
    \Tilde{\rho}_\nu(\chi, \bold k) + \int e^{-i \bold k \cdot \bold u (\chi-\chi')} \pdv{f^0_\nu}{\bold u} d^3u \cdot \int^\chi_0 4\pi iGa^4  \frac{\bold k}{k^2}  \bar \rho_t \Tilde{\delta_t}(\chi, \bold k) d\chi' = \int \Tilde{f'_\nu}(0, \bold k, \bold u) e^{-i \bold k \cdot \bold u \chi} d^3u.
\end{equation}
Using integration by parts and treating the perturbation as first order:
\begin{align}
\label{eq:a12}
    \int e^{-i \bold k \cdot \bold u (\chi-\chi')} \pdv{f^0_\nu}{\bold u} d^3u &= i\bold k(\chi-\chi') \int e^{-i \bold k \cdot \bold u (\chi-\chi')} {f_\nu^0 d^3u}, \\
\label{eq:a13}
    \Tilde{f'_\nu}(0, \bold k, \bold u) &\approx f_\nu^0(0, \bold u) \Tilde{\delta}_\nu(0, \bold k).
\end{align}
Putting Eq.(\ref{eq:a12}) and Eq.(\ref{eq:a13}) into Eq.(\ref{eq:a11}) and defining:
\begin{equation}
    \Phi(\bold q) \equiv \frac{\int f^0_\nu e^{-i \bold q \cdot \bold u d^3u}}{\int f^0_\nu d^3u},
\end{equation}
we have the equation governing the neutrino linear growth:
\begin{equation}
    \Tilde{\delta}_\nu (\chi,\mathbf{k}) = \Phi(\mathbf{k}\chi) \Tilde{\delta}_\nu(0,\mathbf{k}) + 
    4\pi G \int^\chi_0 a^4(\chi')(\chi-\chi') \Phi[\mathbf{k}(\chi-\chi')]
    [\Bar{\rho}_{cdm}(\chi') \Tilde{\delta}_{cdm} (\chi',\mathbf{k}) + \Bar{\rho}_\nu(\chi') \Tilde{\delta}_\nu (\chi',\mathbf{k})] d\chi'.
\end{equation}

We now turn our focus to $\Phi(\bold q)$. The denominator of $\Phi(\bold q)$ can be evaluated numerically. However, the numerator is highly oscillatory:
\begin{equation}
\label{eq:a16}
    I =\int f^0_\nu e^{-i \bold q \cdot \bold u d^3u} = 2\pi \int^\infty_0\int^\pi_0 \frac{u^2 [\cos(qu \cos\theta)-i \sin(qu\cos \theta)]\sin\theta}{e^{mu/T-\xi} +1} du d\theta + \mathrm{anti.},
\end{equation}
where anti. is the contribution from anti-neutrinos, which has $e^{mu/T+\xi} +1$ as the denominator. The imaginary part of the integrand in the right hand side of Eq.(\ref{eq:a16}) vanishes as we integrate it over $\theta$, and the integral $I$ becomes:
\begin{align}
    I &= 2\pi \int^\infty_0 \frac{2u^2 \sin(qu)}{qu(e^{mu/T-\xi} +1)} du + \mathrm{anti.} \nonumber ,\\
\label{eq:a17}
    &=4\pi \frac{T^2}{qm^2} \int^\infty_0 \left[\frac{x \sin(Ax)}{e^{x-\xi} +1 } + \frac{x \sin(Ax)}{e^{x+\xi} +1 }\right] dx,
\end{align}
where $x=mu/T$ and $A=qT/m$. We can expand the anti-neutrino term as a geometric series:
\begin{equation}
    \frac{1}{e^{x+\xi}+1}= \frac{e^{-x-\xi}}{1-(-e^{-x-\xi})} = e^{-(x+\xi)} \sum_{n=0}^\infty (-1)^n e^{-n(x+\xi)} = \sum_{n=1}^\infty (-1)^{n+1} e^{-n(x+\xi)}.
\end{equation}
The integral for anti-neutrino in Eq.(\ref{eq:a17}) becomes:
\begin{equation}
    \int^\infty_0 \frac{x \sin(Ax)}{e^{x+\xi} +1 } dx = \sum_{n=1}^\infty (-1)^{n+1} e^{-n\xi} \frac{2nA}{(A^2+n^2)^2},
\end{equation}
For the neutrino part, we separate the integral into two parts:
\begin{equation}
    \int^\infty_0 \frac{x \sin(Ax)}{e^{x-\xi} +1 } dx = \int^\xi_0 \frac{x \sin(Ax)}{e^{x-\xi} +1 } dx + \int^\infty_0 \frac{y+\xi \sin[A(y+\xi)]}{e^{y} +1 } dy.
\end{equation}
The first term can be evaluated directly, and we expand the second term again. We define:
\begin{align}
    B_1(n) &\equiv \int^\infty_0 e^{-ny} \cos(Ay) dy = \frac{n}{A^2+n^2}\nonumber, \\
    B_2(n) &\equiv \int^\infty_0 e^{-ny} \sin(Ay) dy = \frac{A}{A^2 + n^2}\nonumber, \\
    B_3(n) &\equiv \int^\infty_0 ye^{-ny} \cos(Ay) dy = \frac{n^2-A^2}{(A^2+n^2)^2} , \\
    B_4(n) &\equiv \int^\infty_0 ye^{-ny} \sin(Ay) dy = \frac{2nA}{(A^2+n^2)^2}\nonumber.
\end{align}
Therefore the numerator $I$ is,
\begin{equation}
\begin{split}
    &I = 4\pi \frac{T^2}{qm^2}\int^\xi_0 \frac{x \sin(Ax)}{e^{x-\xi} +1 } dx + 4\pi \frac{T^2}{qm^2} \times \\ 
    \sum_{n=1}^\infty (-1)^{n+1} \{ \xi B_1(n)\sin(A\xi)+&\xi B_2(n)\cos(A\xi)+B_3(n)\sin(A\xi)+B_4(n)\cos(A\xi)[\cos(A\xi)+e^{-n\xi}\},
\end{split}
\end{equation}
and this is how we evaluate $\Phi(\bold q)$ numerically.

%% file: Maintext/posterior.tex
\section{Neutrinos' effects on cosmological parameters}
\label{app:posterior}
We did not separate neutrinos' free-streaming effect from that of the CMB refitting in the N-body simulation due to the computational cost. However, we can extract the neutrinos' effect on the cosmological parameters alone, and we determine the rotation angle between the $(M_\nu, \eta^2)$ and $(x_1, x_2)$ bases. Our regression models are:
\begin{align}
\label{eq:b1}
    \Delta X(M_\nu, \eta^2)[\%] &\equiv \frac{X(M_\nu, \eta^2)}{X(0.06\,\mathrm{eV}, 0)} -1,\nonumber \\
    \Delta X(M_\nu, \eta^2)[\%] &= E_m  M' + E_\eta\eta^2, \\
    \Delta X(x_1, x_2)[\%]&= E_1 x_1 + E_2 x_2\nonumber,
\end{align}
where $X\in\{H_0, \Omega_b, \Omega_c, A_s, n_s\}$, with eigenbasis $x_1=M' \cos \rho\, -\, \eta^2 \sin \rho$ and $x_2 = M' \sin \rho \,+\, \eta^2 \cos \rho$ following similar definitions as in Eq.(\ref{eq:rot}).

\begin{table}[!h]
		\begin{center}
		\begin{tabular}{|c|c c|c|c c|}
			\hline
			Parameter & $E_m$ & $E_\eta$ & $\rho (\pi)$ & $E_1$ & $E_2$\\
			\hline \hline
			$\Delta H_0$ & $-1.62\pm0.02$ & $4.47\pm0.05$ & 0.40 & $-4.76\pm0.05$ & $-0.16 \pm 0.02$ \\
			$\Delta \Omega_b$ & $3.05\pm0.01$ & $-7.06\pm0.03$ & 0.39 & $7.68\pm0.03$ & $0.48\pm0.01$ \\
			$\Delta \Omega_c$ & $3.56\pm0.02$ & $-3.58\pm 0.04$ & 0.39 & $4.58\pm0.04$ & $2.14\pm0.02$ \\
			$\Delta A_s$ & $0.10\pm0.01$ & $2.26\pm0.02$ & 0.40 & $-2.12\pm0.02$ & $0.80\pm0.01$ \\
			$\Delta n_s$ & $-0.10\pm0.00$ & $1.77\pm0.01$ & 0.40 & $-1.71\pm0.01$ & $0.45\pm0.00$ \\
			\hline
		\end{tabular}
		\caption{\label{tab:CP_reg} Regression results for changes of cosmological parameters  as functions of ($M'$, $\eta^2$) and ($x_1, x_2)$ (see Eq.(\ref{eq:b1})).}
		\end{center}
\end{table}

\begin{figure}[!h]
		\begin{center}
		\includegraphics[width=\columnwidth]{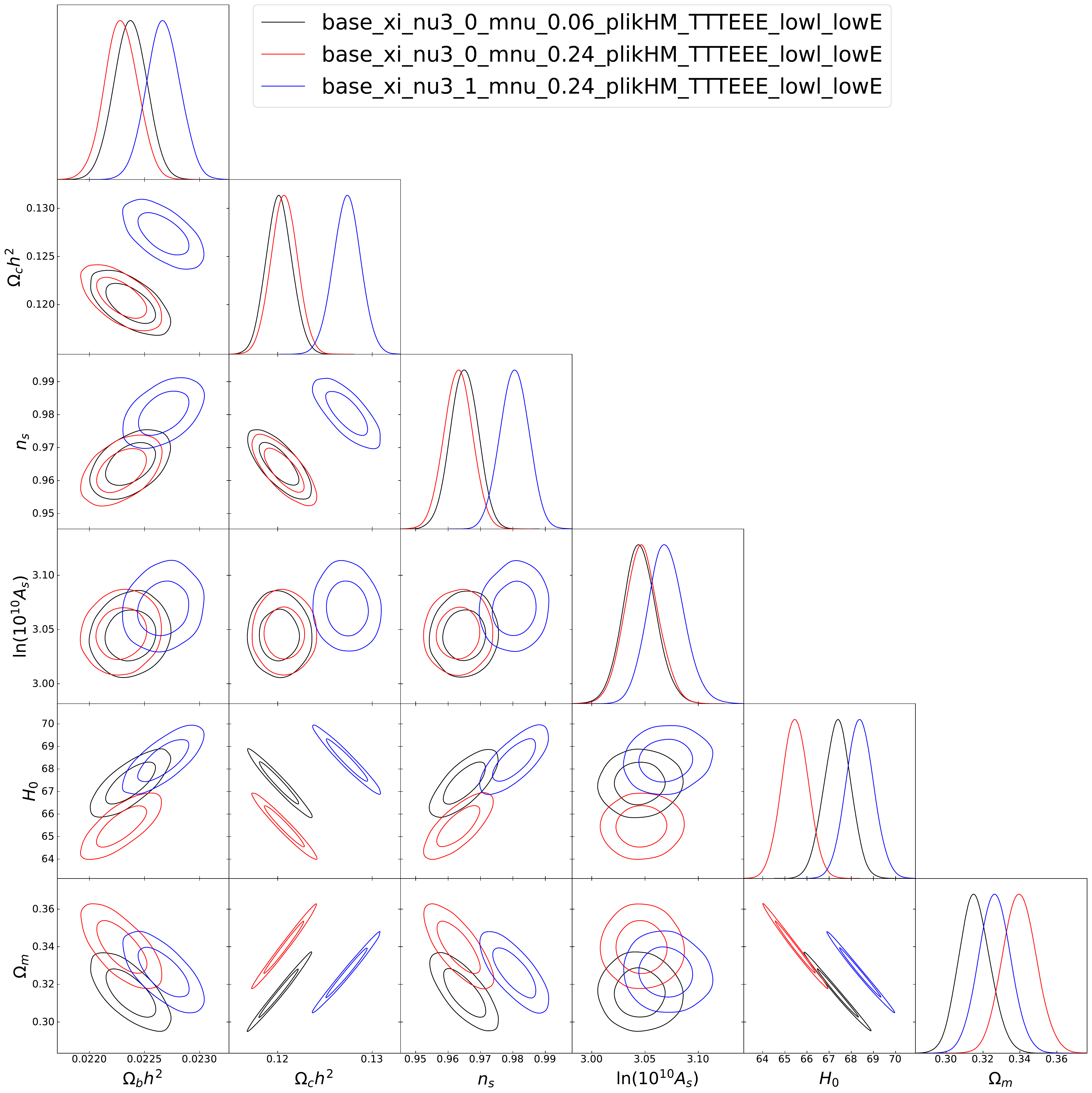}
		\caption{\label{fig:post}1D posterior probability densities and 2D contours ($68\%$ and $95\%$ C.L.) for cosmological parameters extracted from the Planck 2018 CMB data, comparing $M_\nu = 0.06$ eV,  $\eta^2 = 0$ (black lines); $M_\nu = 0.24$ eV,  $\eta^2 = 0$ (red lines); and $M_\nu = 0.24$ eV,  $\eta^2 = 1$ (blue lines)} 
		\end{center}
\end{figure}

From Figure \ref{fig:post} and Table \ref{tab:CP_reg} we can see that the effects of $M_\nu$ and $\eta$ on the cosmological parameters are again opposite to each other. Furthermore, all parameters share the same rotation angle $\rho$ of $0.4\pi$.

